\documentclass[fleqn,twoside]{article}
\usepackage{espcrc2,graphicx,amssymb,cite, epsfig}

\newcommand{\lreal}{L_{10}^r}
\newcommand{\creal}{C_{87}^r}
\newcommand{\leff}{L_{10}^{\rm eff}}
\newcommand{\ceff}{C_{87}^{\rm eff}}
\newcommand{\chpt}{$\chi$PT}
\newcommand{\be}{\begin{equation}}
\newcommand{\ee}{\end{equation}}
\newcommand{\ba}{\begin{eqnarray}}
\newcommand{\ea}{\end{eqnarray}}

\newcommand{\cO}{{\cal O}}
\newcommand{\cJ}{{\cal J}}
\newcommand{\AmS}{{\protect\the\textfont2
  A\kern-.1667em\lower.5ex\hbox{M}\kern-.125emS}}

\title{From Hadronic\ {\Large $\mathbf{\tau}$}\ Decays to the Chiral Couplings $\lreal$ and $\creal$ \thanks{Talk given at the 10th International
Workshop on Tau Lepton Physics (Novosibirsk, 
September 22--25, 2008).}}

\author{Mart\'{\i}n Gonz\'alez-Alonso$^a$\thanks{Speaker. E-mail: Martin.Gonzalez@ific.uv.es.},Antonio Pich$^a$ and Joaquim Prades$^b$\\ [0.4cm]${}^a$Departament de F\'{\i}sica Te\`orica and IFIC, Universitat de Val\`encia-CSIC,\\Apt. de Correus 22085, E-46071 Val\`encia, Spain\\${}^b$ CAFPE and Departamento de F\'{\i}sica Te\'orica y del Cosmos,\\Universidad de Granada, Campus de Fuente Nueva, E-18002 Granada,Spain}
\begin{document}

\begin{abstract}
A sum rule analysis of the hadronic $\tau$-decay data can be used to determine the low-energy constants $\lreal(\mu)$ and $\creal(\mu)$. These constants are QCD chiral-order parameters, which appear at order $p^4$ and $p^6$, respectively, in the chiral perturbation theory expansion of the $V-A$ correlator. At order $p^4$ we obtain
$\lreal(M_\rho) = -(5.22\pm 0.06)\cdot 10^{-3}$.
Including in the analysis the order $p^6$ contributions, we get
$\lreal(M_\rho) = -(4.06\pm 0.39)\cdot 10^{-3}$ and
$\creal(M_\rho) = (4.89\pm 0.19)\cdot 10^{-3}\;\mathrm{GeV}^{-2}$. \vspace{1pc}
\end{abstract}

\maketitle

\section{Introduction}

The fact that the $\tau$ is the only lepton massive enough to decay into hadrons makes it an excellent tool to study QCD, both perturbative and non-perturbative. The main idea under these studies is the use of the analyticity properties of the different two-point correlation functions, which are relevant for the dinamical description of the $\tau$ hadronic width.

As it is well known, analyticity of two-point correlation functions allow us to relate different regions of the $q^2$-complex plane. Roughly speaking, one can relate in this way regions where we are able to compute analytically, either with Chiral Perturbation Theory (\chpt) or with the short-distance Operator Product Expansion (OPE), with regions of the complex plane where we are not able to compute (except in the lattice) but that are experimentally accessible. This connection can be used either to predict observables that we are not able to calculate \lq\lq directly'' or, in the other way around, to extract the value of QCD parameters that are not fixed theoretically.

An excellent example of this last strategy is the accurate determination of the QCD coupling $\alpha_s(M_\tau)$ \cite{alphas,LDP:92,DHZ06,new08,review} from the inclusive $\tau$ decay width into hadrons, which becomes the most precise determination of $\alpha_s(M_Z)$ after QCD running. Other parameters of the Standard Model that have been extracted from $\tau$ physics are the strange quark mass and the Cabibbo-Kobayashi-Maskawa quark-mixing $|V_{us}|$ \cite{su3,CKP98,KKP01,MALT,Vus}.

Non-perturbative QCD quantities can also be obtained from $\tau$-decay data. The fact that the spectral function of the $\tau$ decay can be separated experimentally in its vector and axial-vector contributions allows us to study their difference, that is specially interesting because it vanishes in perturbative QCD (in the chiral limit) and therefore it is a purely non-perturbative quantity.

The $\tau$-decay measurement of this $V-A$ spectral function has been used to perform \cite{DG:94,DHG98,NAR01} phenomenological tests of the so-called Weinberg sum rules (WSRs) \cite{WSR}, to compute the electromagnetic mass difference between the charged and neutral pions \cite{DHG98}, to determine several QCD vacuum condensates \cite{DS07,CGM03} and also to determine the $\Delta I=3/2$ contribution of the $\Delta S=1$ four-quark operators $Q_7$ and $Q_8$ to $\varepsilon_K'/\varepsilon_K$, in the chiral limit \cite{Q7Q8}.

Using \chpt\ \cite{WEI79,GL84,GL85}, the hadronic $\tau$-decay data can also be related to order parameters of the spontaneous chiral symmetry breaking (S$\chi$SB) of QCD. \chpt\ is the effective field theory of QCD at very low energies that describes the physics of the S$\chi$SB Nambu-Goldstone bosons through an expansion in external momenta and quark masses, with coefficients that are order parameters of S$\chi$SB. At lowest order (LO), i.e. $\cO(p^2)$, all low-energy observables are described in terms of the pion decay constant $f_\pi \simeq 92.4$ MeV and the light quark condensate. At $\cO(p^4)$, the SU(3) \chpt\ Lagrangian contains 12 low-energy constants (LECs), $L_{i=1,\cdots,10}$ and $H_{1,2}$ \cite{GL85}, whereas at $\cO(p^6)$ we have 94 \, (23) additional parameters $C_{i=1,\cdots,94} \, (C^W_{i=1,\cdots,23})$ in the even (odd) intrinsic parity sector \cite{p6}. These LECs are not fixed by symmetry requirements alone and have to be determined phenomenologically or using non-perturbative techniques. The $L_i$ couplings have been determined in the past to an acceptable accuracy (a recent compilation can be found in ref.~\cite{ECK07}), but much less well determined are the ${\cal O}(p^6)$ couplings $C_i$.

There has been a lot of recent activity to determine the chiral LECs from theory, using as much as possible QCD information \cite{MOU97,KN01,RPP03,BGL03,CEE04,CEE05,KM06,RSP07,MP08,PRS08}. This strong effort is motivated by the precision required in present phenomenological applications, which makes necessary to include corrections of $\cO(p^6)$ where the huge number of unknown couplings are the major source of theoretical uncertainty.

Here we present an accurate determination of the \chpt\ couplings $L_{10}$ and $C_{87}$ \cite{GPP08}, using the most recent experimental data on hadronic $\tau$ decays \cite{ALEPH05}. Previous work on $L_{10}$ using $\tau$-decay data can be found in refs.~\cite{DHG98,NAR01,DS07,DS04}. Our analysis is the first one which includes the known two-loop \chpt\ contributions and, therefore, provides also the $\cO(p^6)$ coupling $C_{87}$.

We will first introduce the sum rule relations that we will use, then we will show our results and finally we will compare them with other recent analytic results and hadronic $\tau$ data determinations.

\section{Theoretical framework: the sum rules}

\begin{figure}[tbh]
\centering
\includegraphics[width=0.4\textwidth]{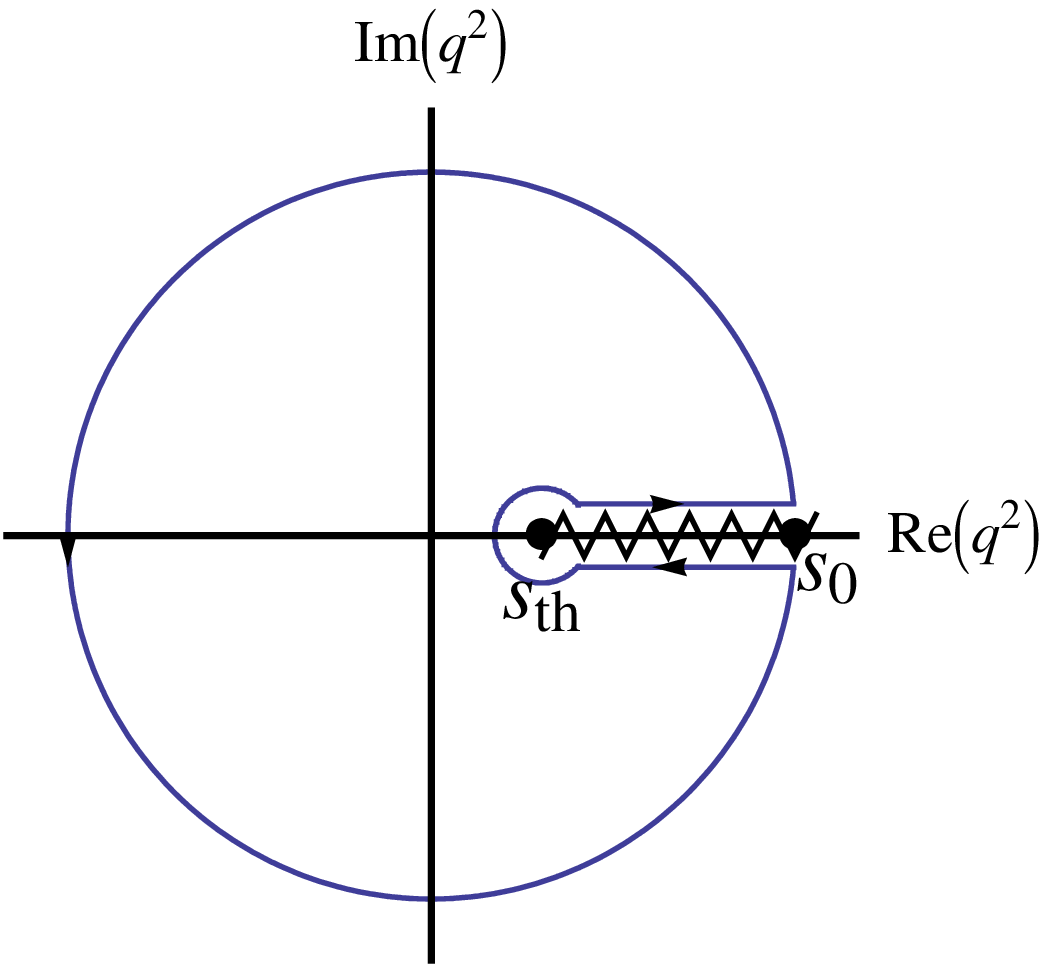}
\caption[]{Analytic structure of $\overline{\Pi}(s)$.}
\label{fig:circuit}
\end{figure}

The basic objects of the theoretical analysis are two-point correlation functions of the non-strange vector and axial-vector quark currents
\ba
\label{eq:two}
\Pi^{\mu\nu}_{ij,\cJ}(q)\equiv i \!\int\!\! \mathrm{d}^4 x \,  \mathrm{e}^{i q x}\langle 0 | T \!\left( \cJ_{ij}^\mu(x) \cJ_{ij}^\nu(0)^\dagger \right)\!| 0 \rangle \nonumber \\
\!\!\!\!=\! (-\!g^{\mu\nu}\! q^2 \!+\! q^\mu\! q^\nu ) \Pi^{(1)}_{ij,\cJ}(q^2) + q^\mu\! q^\nu \Pi^{(0)}_{ij,\cJ}(q^2) ,\!\!
\ea
where $\cJ_{ij}^\mu$ denotes $V_{ud}^\mu\!=\!\overline{u} \gamma^\mu d$ and $A_{ud}^\mu\!=\!\overline{u} \gamma^\mu \gamma_5 d$. In particular, we are interested in the difference $\Pi(s) \equiv \Pi_{ud,V}^{(0+1)}-\Pi_{ud,A}^{(0+1)}$, and we will work in the isospin limit ($m_u=m_d$) where $\Pi^{(0)}_{ud,V}(q^2)=0$. The analytic behaviour of this correlator is shown in Fig.\ref{fig:circuit}, together with the complex circuit that we will use to apply the Cauchy's theorem.

As we are interested in relating the \chpt\ domain (very low energies) with the $\tau$ data, we multiply this correlator by a weight function of the form $1/s^n$ (with $n>0$). In this way we generate a residue at $s=0$ when we integrate over the circuit of Fig. \ref{fig:circuit} and apply Cauchy's theorem. Taking into account the OPE associated with our correlator at large momenta and working in
particular with the cases $n=1,2$, one gets the following sum rules (see ref.~\cite{GPP08} for a careful derivation):

\ba
\label{eq:SR1}
-8 \, L_{10}^{\rm eff} \!\!\!\!&\equiv&\!\!\!\! \int^{\infty}_{s_{\rm th}} \frac{\mathrm{d}s}{s} \frac{1}{\pi} \, {\rm Im} \, \Pi(s)= \frac{2 f_\pi^2}{m_\pi^{2}} +\! \Pi(0)~, \\
\label{eq:SR2}16
\, C_{87}^{\rm eff} \!\!\!\!&\equiv&\!\!\!\! \int^{\infty}_{s_{\rm th}} \frac{\mathrm{d}s}{s^2} \frac{1}{\pi} \, {\rm Im} \, \Pi(s)= \frac{2 f_\pi^2}{m_\pi^{4}} +\! \frac{\rm{d}\Pi}{\rm{ds}}(0)~,
\ea
where the integrations start at the threshold $s_{\rm th}=4 m_\pi^2$.
These two relations represent the starting point of our work and define the effective parameters $\leff$ and $\ceff$. Their interest stems from the fact that the l.h.s. can be extracted from the data (see Section \ref{sec:eff}), while the r.h.s. can be rigourously calculated within \chpt~in terms of the LECs that we want to determine (see Section \ref{sec:real}).

\section{The $\tau$-data side}
\label{sec:eff}
We will use the recent ALEPH data on hadronic $\tau$ decays \cite{ALEPH05}, that provide the most precise measurement of the $V-A$ spectral function. In the integrals of equations (\ref{eq:SR1}) and (\ref{eq:SR2}) we are forced to cut the integration at a finite value $s_0$, neglecting in this way the rest of the integral from $s_0$ to infinity. The superconvergence properties of $\Pi(s)$ at large momenta imply a tiny contribution from the neglected range of integration, provided $s_0$ is large enough. Nevertheless, this
generates a theoretical error called quark-hadron duality violation (DV)\footnote{Equivalently, we are assuming that the OPE is a good approximation for $\Pi(s)$ at any $|s|\!\!=\!\!s_0$, what is not expected to happen near the real axis, and that produces the DV.}. From the $s_0$-sensitivity of the effective parameters one can assess the size of this error.

\begin{figure}[tbh]
\vfill
\centerline{\begin{minipage}[t]{.3\linewidth}
\centering
\centerline{\includegraphics[width=7cm]{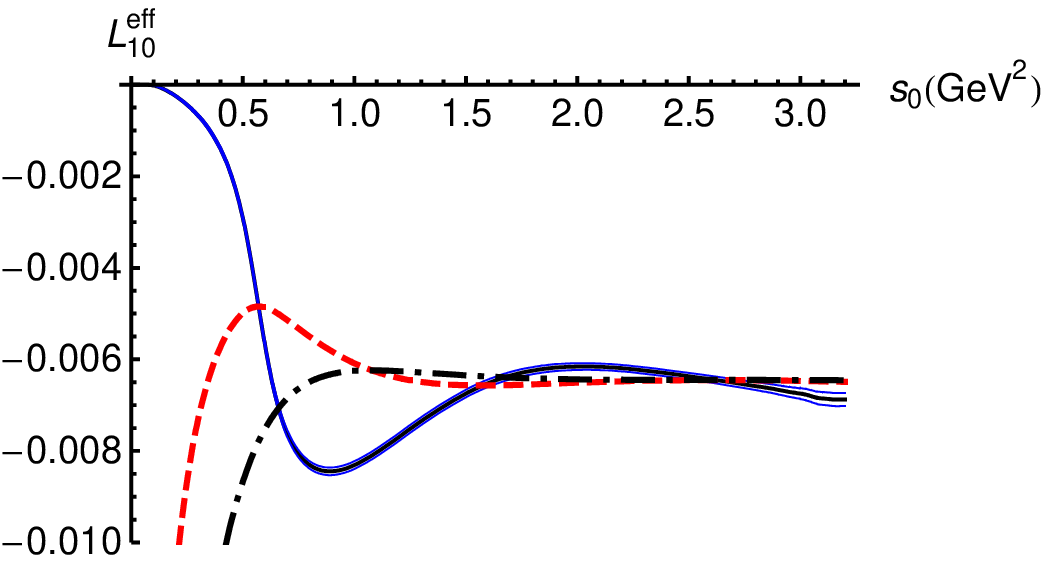}}
\end{minipage}}
\vspace{0.67cm}
\centerline{\begin{minipage}[t]{.3\linewidth}
\centering
\centerline{\includegraphics[width=7cm]{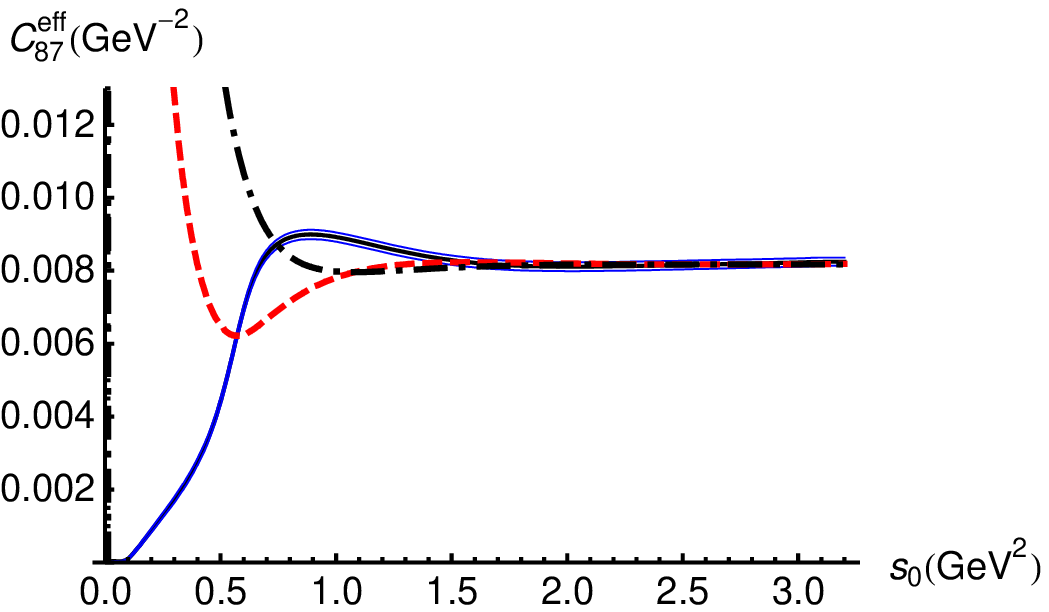}}
\end{minipage}}
\vfill
\caption{$L_{10}^{\rm eff}(s_0)$ and $C_{87}^{\rm eff}(s_0)$ from different sum rules. For clarity, we do not include the error bands associated with the modified weights.}
\label{fig:L10C87}
\end{figure}

In Fig. \ref{fig:L10C87}, we plot the value of $L_{10}^{\rm eff}$ obtained for different values of $s_0$, with the one-sigma experimental error band, and we can see a quite stable result at $s_0\!\gtrsim\! 2~\mathrm{GeV}^2$
(solid lines). The weight function $1/s$ decreases the impact of the high-energy region, minimising the DV; the resulting integral appears then to be much better behaved than the sum rules with $s^n$ ($n\ge0$) weights.

There are some possible strategies to estimate the value of $L_{10}^{\rm eff}$ and his error. One is to give the predictions fixing $s_0$ at the so-called ``duality points'', two points where the first and second 
WSRs \cite{WSR} happen to be satisfied. In this way we get $L_{10}^{\rm eff} = -(6.50\pm 0.13) \cdot 10^{-3}$, where the uncertainty covers the values obtained at the two ``duality points''. If we assume that the integral (\ref{eq:SR1}) oscillates around his asymptotic value with decreasing oscillations and we perform an average between the maxima and minima of the oscillations we get $L_{10}^{\rm eff} = -(6.5\pm 0.2) \cdot 10^{-3}$. Another way of estimating the DV uses appropriate oscillating functions defined in \cite{GON07} which mimic the real quark-hadron oscillations above the data. These functions are defined such that they match the data at $\sim\!3~\mathrm{GeV}^2$, go to zero with decreasing oscillations and satisfy the two WSRs. We find in this way $L_{10}^{\rm eff} = -(6.50\pm 0.12) \cdot 10^{-3}$, where the error spans the range generated by the different functions used. Finally we can take advantage of the WSRs to construct modified sum rules with weight factors $w(s)$ proportional to $(1-s/s_0)$, in order to suppress numerically the role of the suspect region around $s\!\sim\!s_0$ \cite{LDP:92}. Fig.~\ref{fig:L10C87} shows the results obtained with $w_1(s)\!\equiv\!\left(1\!-\!s/s_0\right)/s$ (dashed line) and $w_2(s)\!\equiv\!\left(1\!-\!s/s_0\right)^2/s$ (dot-dashed line). These weights give rise to very stable results over a quite wide range of $s_0$ values. One gets $L_{10}^{\rm eff} = -(6.51\pm 0.06) \cdot 10^{-3}$ using $w_1(s)$ and $L_{10}^{\rm eff} = -(6.45\pm 0.06) \cdot 10^{-3}$ using $w_2(s)$. Taking into account all the previous discussion, we quote as our final result:
\be
\label{L10eff}
L_{10}^{\rm eff} = -(6.48\pm 0.06) \cdot 10^{-3} \, .
\ee

We have made a completely analogous analysis to determine $C_{87}^{\rm eff}$. The results are shown in Fig.~\ref{fig:L10C87}. The solid lines, obtained from Eq.~(\ref{eq:SR2}), are much more stable than the corresponding results for $L_{10}^{\rm eff}$, due to the $1/s^2$ factor in the integrand. The dashed and dot-dashed lines have been obtained with the modified weights $\rm{w_3(s)\equiv\! \frac{1}{s^2}\!\left(1\!-\!\frac{s^2}{s_0^2}\right)}$ and $\rm{w_4(s)\equiv\! \frac{1}{s^2} \left(1\!-\!\frac{s}{s_0}\right)^2 \left(1\!+\!2\frac{s}{s_0}\right)}$. The agreement among the different estimates is quite remarkable, and our final result is
\be
\label{C87eff}
C_{87}^{\rm eff} = (8.18\pm 0.14) \cdot 10^{-3} \, {\rm GeV}^{-2} \, .
\ee
\section{The \chpt\ side}
\label{sec:real}
Using the results of ref.~\cite{ABT00} to calculate within \chpt\ the l.h.s. of equations (\ref{eq:SR1}) and (\ref{eq:SR2}) we get
\ba
\label{L10-p6}
-8~\leff &\!\! =&\!\! -~8~L_{10}^r(\mu) + G^4_{1L}(\mu)  \nonumber\\
&&\!\! +~G^6_{0L}(\mu) + G^6_{1L}(\mu) + G^6_{2L}(\mu) ~\!, \\
\label{C87-p6}
16~\ceff &\!\! =&\!\! H^4_{1L} \nonumber\\
&&\!\! +~\!16~\!C_{87}^r(\mu)+H^6_{1L}\!(\mu)+H^6_{2L}\!(\mu) ~\!,
\ea
where the functions $G^m_{nL}(\mu), H^{m}_{nL}(\mu)$ are corrections of order $p^m$ generated at the $n$-loop level, which explicit analytic form \cite{GPP08} is omitted for simplicity.

Working at $\cO(p^4)$ the determination is straightforward and one gets
\be
\label{valL10p4}
L_{10}^r(\mu\!=\!M_\rho) = -(5.22 \pm 0.06) \cdot 10^{-3}  \, .
\ee
At order $p^6$, the numerical relation is more involved because the small corrections $G^6_{0L,1L}(\mu)$ contain some LECs that represent the main source of uncertainty for $L_{10}^r$. It is useful to classify the $\cO(p^6)$ contributions through their ordering within the $1/N_C$ expansion. The tree-level term $G_{0L}^6(\mu)$ contains the only $\cO(p^6)$ correction in the large--$N_C$ limit, $\rm{4 m_\pi^2 (C_{61}^r \!-\! C_{12}^r \!-\! C_{80}^r)}$; this correction is numerically small because of the $m_\pi^2$ suppression and can be estimated with a moderate accuracy \cite{KM06,JOP04,UP08,CEE05,ABT00}. At NLO $G_{0L}^6(\mu)$ contributes with a term of the form $m_K^2(C_{62}^r \!-\! C_{13}^r \!-\! C_{81}^r)$. In the absence of information about these LECs we will adopt the conservative range $|C_{62}^r \!- C_{13}^r \!- C_{81}^r\!| \le |C_{61}^r \!- C_{12}^r \!- C_{80}^r|/3$, which generates the uncertainty that will dominate our final error on $L_{10}^r$. Also at this order in $1/N_C$ there is the one-loop correction $G_{1L}^6(\mu)$, that is proportional to $L_{9}^r$ which is better known \cite{BT02}.
Calculating the $1/N_C^2$ suppressed two-loop function $G_{2L}^6(\mu)$ and taking all these contributions into account we finally get the wanted $\cO(p^6)$ result:
\ba
\label{valL10p6}
L_{10}^r(M_\rho) \!\!\!&=&\!\!\! -(4.06 \pm 0.04_{L_{10}^{\mathrm{eff}}}\pm 0.39_{\mathrm{LECs}}) \cdot 10^{-3} \nonumber\\
\!\!\!&=&\!\!\! -(4.06 \pm 0.39) \cdot 10^{-3} \, .
\ea
where the error has been split into its two main components.
Repeating the same process with $C_{87}^r$ (where the only LEC involved is $L_9^r$) we get
\be
\label{valC87}
C_{87}^r(M_\rho) = (4.89 \pm 0.19) \cdot 10^{-3} \:\mathrm{GeV}^{-2}\, .
\ee
\section{Summary}
Through a sum rule analysis, that only uses general properties of QCD and the measured $V\!-\!A$ spectral function \cite{ALEPH05}, we have determined the chiral LECs $L_{10}^r(M_\rho)$ and $C_{87}^r(M_\rho)$ rather accurately, with a careful analysis of the theoretical uncertainties.

There are other determinations of $L_{10}$ from $\tau$ data in the literature. Our result for $L_{10}^{\rm eff}$ agrees with \cite{DHG98,DS04,DS07}, but our estimation includes a more careful assessment of the theoretical errors. The $\rm{3.2\,\sigma}$ discrepancy between the estimation of ref.~\cite{NAR01} and ours is caused by an underestimation of the systematic error associated with the duality-point approach used in that reference. In ref.~\cite{DS04} $C_{87}^{\rm eff}$ is also determined, in good agreement with our result. The extraction of $L_{10}^r(\mu)$ from $L_{10}^{\rm eff}$ has only been done previously in ref.~\cite{DHG98}, at $\cO(p^4)$.

Our determinations of $L_{10}^r(M_\rho)$ and $C_{87}^r(M_\rho)$ agree within errors with the large--$N_C$ estimates based on lowest-meson dominance \cite{KN01,CEE04,ABT00,PI02}, $L_{10} \approx -3 f_\pi^2 / (8 M_V^2) \approx -5.4\cdot 10^{-3}$ and $C_{87} \approx 7 f_\pi^2 / (32 M_V^4) \approx 5.3\cdot 10^{-3}~\mathrm{GeV}^{-2}$, and with the result of ref. \cite{MP08} for $C_{87}$, based on Pad\'e approximants. These predictions, however, are unable to fix the scale dependence which is of higher-order in $1/N_C$. More recently, the resonance chiral theory Lagrangian \cite{CEE04,EGPdR89} has been used to analyse the correlator $\Pi(s)$ at NLO order in the $1/N_C$ expansion. Matching the effective field theory description with the short-distance QCD behaviour, the two LECs are determined, keeping full control of their $\mu$ dependence. The theoretically predicted values $L_{10}^r(M_\rho) = -(4.4 \pm 0.9) \cdot 10^{-3}$ and $C_{87}^r(M_\rho)=(3.6 \pm 1.3) \cdot 10^{-3}$  GeV$^{-2}$ \cite{PRS08} are in perfect agreement with our determinations, although less precise. A recent lattice estimate \cite{LAT08} finds $L_{10}^r(M_\rho) = -(5.2 \pm 0.5) \cdot 10^{-3}$ at order $p^4$, in good agreement with our result (\ref{valL10p4}).

Using the results of ref.~\cite{GHI07}, the SU(2) $\chi PT$ LEC $\overline l_5$ can be extracted from $L_{10}^r(\mu)$. We find $\rm{\overline l_5 = 13.30 \pm 0.11}$ at $\cO(p^4)$ and $\overline l_5 = 12.24 \pm 0.21$ at $\cO(p^6)$.

Recent analyses of the decay $\pi^+ \!\to\! l^+ \nu \gamma$ at $\cO(p^6)$ have provided accurate values for the combinations $L_9\!+\!L_{10}$ \cite{UP08} and $\overline l_5\! -\! \overline l_6$ \cite{BT97}, that can be combined with our results to get $L_9^r(M_\rho)=(5.5 \pm 0.4) \cdot 10^{-3}$ and $\overline l_6 = 15.22 \pm 0.39$ to order $p^6$, in perfect agreement with refs.~\cite{BT02,BCT98}.

\section*{Acknowledgements}M. G.-A. is indebted to MICINN (Spain) for an FPU Grant. Work partly supported by the EU network FLAVIAnet [Contract No MRTN-CT-2006-035482], by MICINN, Spain [Grants No FPA2007-60323, No FPA2006-05294 and No CSD2007-00042 --CPAN--], by Junta de Andaluc\'{\i}a [Grants No P05-FQM 191, No P05-FQM 467 and No P07-FQM 03048] and by Generalitat Valenciana [Grant No Prometeo/2008/069].


\begin{thebibliography}{9}

\bibitem{alphas}
E. Braaten, Phys. Rev. Lett. {\bf 60} (1988) 1606;
Phys. Rev. {\bf D 39} (1989) 1458;
S. Narison and A. Pich, Phys. Lett.
{\bf B 211} (1988) 183;
E. Braaten, S. Narison and A. Pich,
Nucl. Phys. {\bf B 373} (1992) 581;
F. Le Diberder and A. Pich,
Phys. Lett. {\bf B 286} (1992) 147;
A. Pich, Nucl. Phys. B (Proc. Suppl.)
{\bf 39 B,C} (1995) 326.

\bibitem{LDP:92}
F. Le Diberder and A. Pich, Phys. Lett. {\bf B 289} (1992) 165.

\bibitem{DHZ06} M. Davier, A. H\"ocker and H. Zhang,
Rev. Mod. Phys. {\bf 78} (2006) 1043;
M. Davier {\it et al.}, 
Eur. Phys. J. {\bf C 56} (2008) 305.

\bibitem{new08}
P.A. Baikov, K.G. Chetyrkin and J.H. K\"uhn, 
 Phys. Rev. Lett.  {\bf 101} (2008) 012002;
M. Beneke and M. Jamin, JHEP {\bf 09} (2008) 044;
K.~Maltman and T. Yavin, 
Phys. Rev. {\bf D 78} (2008) 094020. 

\bibitem{review}
A. Pich, Int. J. Mod. Phys. {\bf A 21}
(2006) 5652;
Nucl. Phys. B (Proc. Suppl.) {\bf 169} (2007) 393;
arXiv:0806.2793 [hep-ph].

\bibitem{su3}
A. Pich and J. Prades,
JHEP {\bf 06} (1998) 013;
Nucl. Phys. B (Proc. Suppl.)  {\bf 74} (1999) 309;
JHEP {\bf 10} (1999) 004;
Nucl. Phys. B (Proc. Suppl.)  {\bf 86} (2000) 236;
J. Prades, Nucl. Phys. B (Proc. Suppl.)  {\bf 76} (1999) 341;
S. Chen, {\it et al.}
  Eur. Phys. J.  {\bf C 22} (2001) 31;
M. Davier, {\it et al.}
  Nucl. Phys. B (Proc. Suppl.)  {\bf 98} (2001) 319;

\bibitem{CKP98}
K.G. Chetyrkin, J.H. K\"uhn and A.A. Pivovarov,
Nucl. Phys. {\bf B 533} (1998) 473;
P.A. Baikov, K.G. Chetyrkin and J.H. K\"uhn,
Phys. Rev. Lett. {\bf 95} (2005) 012003.

\bibitem{KKP01}
J.G. K\"orner, F. Krajewski and A.A. Pivovarov,
Eur. Phys. J {\bf C 20} (2001) 259.

\bibitem{MALT}
K. Maltman, Phys. Rev. {\bf D 58} (1998) 093015;
J. Kambor and K. Maltman, ibid. {\bf D 62}
(2000) 093023;
ibid.  {\bf D 64} (2001) 093014;
K. Maltman and C.E. Wolfe, Phys. Lett.
{\bf B 639} (2006) 286;
K. Maltman {\it et al.}, arXiv:0807.3195 [hep-ph].

\bibitem{Vus}
E. G\'amiz {\it et al.} JHEP {\bf 01} (2003) 060;
  Phys. Rev. Lett.  {\bf 94} (2005) 011803;
 Nucl. Phys. B (Proc. Suppl.)  {\bf 144} (2005) 59;
arXiv:hep-ph/0505122;
arXiv:hep-ph/0610246;
Nucl. Phys. B (Proc. Suppl.)  {\bf 169} (2007) 85;
PoS {\bf KAON} (2008) 008.

\bibitem{DG:94}
J. F. Donoghue and E. Golowich, Phys. Rev. {\bf D 49} (1994) 1513.

\bibitem{DHG98} M. Davier, A. H\"ocker, L. Girlanda,
and J. Stern, Phys. Rev.  {\bf D 58} (1998) 096014.

\bibitem{NAR01} S.~Narison,
  Nucl. Phys.  B {\bf 593} (2001) 3.

\bibitem{WSR} S. Weinberg, Phys. Rev. Lett. {\bf 18} (1967) 507.

\bibitem{DS07}
C.A. Dom\'{\i}nguez and K. Schilcher,
  JHEP {\bf 01} (2007) 093;
  J. Bordes, C.A. Dom\'{\i}nguez, J. Pe\~narrocha and K. Schilcher,
  JHEP {\bf 02} (2006) 037.

\bibitem{CGM03}
 V. Cirigliano, E. Golowich and K. Maltman,
  Phys. Rev.  {\bf D 68} (2003) 054013.

\bibitem{Q7Q8}
J.F. Donoghue and E. Golowich, Phys. Lett.
{\bf B 478} (2000) 172;
V. Cirigliano {\it et al.}, ibid. {\bf B 522} (2001) 245;
ibid. {\bf B 555} (2003) 71;
J. Bijnens, E. G\'amiz and J. Prades, JHEP {\bf 10}
(2001) 009;
Nucl. Phys. B (Proc. Suppl.) {\bf 121} (2003) 195;
S. Narison, Nucl. Phys. {\bf B 593} (2001) 3.

\bibitem{WEI79}
S. Weinberg,  Physica {\bf A 96} (1979) 327.

\bibitem{GL84}
J. Gasser and H. Leutwyler,
Annals Phys.  {\bf 158} (1984) 142.

\bibitem{GL85}
J. Gasser and H. Leutwyler,
 Nucl. Phys. {\bf B 250} (1985) 465.

\bibitem{p6}
J. Bijnens, L. Girlanda and P. Talavera, Eur. Phys. J {\bf C23} (2002) 539;
 J. Bijnens, G. Colangelo and G. Ecker,
Annals Phys.  {\bf 280} (2000) 100;
  JHEP {\bf 02} (1999) 020;
H.W. Fearing and S. Scherer,
  Phys. Rev.  {\bf D 53} (1996) 315.

\bibitem{ECK07}
 G. Ecker,   Acta Phys. Polon. {\bf B 38} (2007) 2753.

\bibitem{MOU97}
B. Moussallam, Nucl. Phys. {\bf B 504} (1997) 381.

\bibitem{KN01}
M. Knecht and A. Nyffeler, Eur. Phys. J  {\bf C 21} (2001) 659.

\bibitem{RPP03}
P. Ruiz-Femen\'{\i}a, A. Pich and J. Portol\'es, JHEP {\bf 07} (2003) 003.

\bibitem{BGL03}
J. Bijnens, E. G\'amiz, E. Lipartia and J. Prades,
  JHEP {\bf 04} (2003) 055.

\bibitem{KM06}
K. Kampf and B. Moussallam,
Eur. Phys. J. {\bf C 47} (2006) 723;
S. D\"urr and J. Kambor, Phys. Rev.
{\bf D 61} (2000) 114025.

\bibitem{CEE05} V. Cirigliano {\it et al.},
JHEP {\bf 04} (2005) 006.

\bibitem{CEE04}
V. Cirigliano {\it el al.},
Nucl. Phys. {\bf B 753} (2006) 139;
 Phys. Lett.  B {\bf 596} (2004) 96.

\bibitem{RSP07} I. Rosell, J.J. Sanz-Cillero and A. Pich,
  JHEP {\bf 01} (2007) 039.

\bibitem{MP08}
P. Masjuan and S. Peris,
 Phys.  Lett.  {\bf B 663} (2008) 61;
  JHEP {\bf 05} (2007) 040.

\bibitem{PRS08}
A. Pich, I. Rosell and J.J. Sanz-Cillero,
 JHEP {\bf 07} (2008) 014. 

\bibitem{GPP08} M.~Gonz\'alez-Alonso, A.~Pich and J.~Prades,
Phys. Rev. D, in press (arXiv:0810.0760 [hep-ph]);
Nucl. Phys. B (Proc. Suppl.), in press (arXiv:0810.2459 [hep-ph]).

\bibitem{ALEPH05} S. Schael {\it et al.}
[ALEPH Collaboration], Phys. Rep. {\bf 421} (2005) 191.

 \bibitem{DS04}
 C.A. Dom\'{\i}nguez and K. Schilcher,
  Phys.\ Lett.\  {\bf B 581} (2004) 193;
  ibid. {\bf B 448} (1999) 93.

\bibitem{GON07} M. Gonz\'alez-Alonso,
Master's thesis, Val\`encia Univ (2007).

\bibitem{ABT00} G. Amor\'os, J. Bijnens and P. Talavera,
Nucl. Phys. {\bf B 568} (2000) 319.

\bibitem{JOP04} M. Jamin, J.A. Oller and A. Pich, JHEP {\bf 02} (2004) 047.

\bibitem{UP08}
R. Unterdorfer and H. Pichl,
  Eur.\ Phys.\ J.  {\bf C 55} (2008) 273.

\bibitem{BT02} J. Bijnens and P. Talavera,
JHEP {\bf 03} (2002) 046.

\bibitem{GHI07}
 J. Gasser, C. Haefeli, M.A. Ivanov and M. Schmid,
  Phys.\ Lett.\  B {\bf 652} (2007) 21.

\bibitem{PI02} A. Pich, arXiv:hep-ph/0205030.

\bibitem{EGPdR89}
G. Ecker, J. Gasser, A. Pich and E. de Rafael, Nucl. Phys.
{\bf B 321} (1989) 311;
G. Ecker, J. Gasser, H. Leutwyler, A. Pich and E. de Rafael,
Phys. Lett. {\bf B 223} (1989) 425.

\bibitem{LAT08}
  E.~Shintani {\it et al.}  [JLQCD Collaboration],
Phys. Rev. Lett. {\bf 101} (2008) 242001.

\bibitem{BT97}
J. Bijnens and P. Talavera,
Nucl. Phys. B {\bf 489} (1997) 387.

\bibitem{BCT98}
J.~Bijnens, G.~Colangelo and P.~Talavera,
JHEP {\bf 05} (1998) 014

\end{thebibliography}
\end{document}